\documentclass{emulateapj}
\newcommand{\sax}{SAX~J1808.4$-$3658}
\newcommand{\saxs}{J1808.4}

\begin{document}

\title{Multiband Studies of the Optical Periodic Modulation 
in the X-Ray Binary SAX~J1808.4$-$3658 during Its Quiescence and 2008 Outburst} 

\author{Zhongxiang Wang\altaffilmark{1}, 
Rene P. Breton\altaffilmark{2,3}, 
Craig O. Heinke\altaffilmark{4}, 
Christopher J. Deloye \altaffilmark{4},
and Jing Zhong\altaffilmark{1,5}}

\altaffiltext{1}{\footnotesize Shanghai Astronomical Observatory,
Chinese Academy of Sciences, 80 Nandan Road, Shanghai 200030, China}

\altaffiltext{2}{\footnotesize School of Physics and Astronomy,
University of Southampton, Highfield, 
Southampton SO17 1BJ, U.K.}

\altaffiltext{3}{\footnotesize Department of Astronomy and Astrophysics,
University of Toronto, 50 St. George Street,
Toronto, ON M5S 3H4, Canada}

\altaffiltext{4}{\footnotesize Department of Physics,
University of Alberta, Room 238 CEB,
Edmonton, AB T6G 2G7,
Canada}

\altaffiltext{5}{\footnotesize Graduate School of Chinese Academy of Sciences,
No. 19A, Yuquan Road, Beijing 100049,
China}

\begin{abstract} 
We report on time-resolved optical imaging of the X-ray binary \sax\ 
during its quiescent state and 2008 outburst. 
The binary, containing an accretion-powered millisecond 
pulsar, has a large sinusoidal-like modulation in its quiescent optical 
emission. We employ a Markov chain Monte Carlo technique
to fit our multi-band light curve data in quiescence with 
an irradiated star model, and derive a tight constraint of 
$50^{+6}_{-5}$ deg on the inclination angle $i$ of the binary system.
The pulsar and its companion are constrained to have masses
of $0.97^{+0.31}_{-0.22}\ M_{\sun}$ and $0.04^{+0.02}_{-0.01}\ M_{\sun}$ 
(both 1$\sigma$ ranges), 
respectively. The dependence of these results on the measurements of 
the companion's projected radial velocity is discussed. 
We also find that the accretion disk had nearly constant optical
fluxes over a $\sim$500 day period in the quiescent state our data covered, 
but started brightening 1.5 months before the 2008 
outburst. Variations in modulation during the outburst were detected in our 
four observations made 7-12 days after the start of the outburst, and 
a sinusoidal-like modulation with 0.2 mag amplitude changed to have
a smaller amplitude of 0.1 mag. The modulation variations are discussed. 
We estimate the albedo of the companion during its quiescence and the outburst,
which was approximately 0 and 0.8 (for isotropic emission), respectively.
This large difference probably provides additional evidence that
the neutron star in the binary turns on as a radio pulsar in quiescence.

\end{abstract}


\section{INTRODUCTION}
 
Among $\sim$200 known low-mass X-ray binaries (LMXBs), \sax\ (hereafter \saxs)
stands out because not only was it the first discovered accretion-powered 
millisecond pulsar (APMP) system \citep{wv98}, but also this transient 
system is relatively bright during quiescence and has relatively frequent 
outbursts, 
allowing detailed studies of its various properties (see \citealt{har+08} 
and references therein). One intriguing property of the system 
is the fact that the millisecond pulsar possibly 
switches its energy source from accretion during 
outbursts to rotation during quiescence. This possible property is indicated
by bright, large-amplitude optical modulation 
(\citealt{hom+01}; \citealt{cam+04}; \citealt{del+08}; \citealt{wan+09}) 
seen in the binary during 
quiescence, which can not be caused by X-ray heating of the inner face of
the companion star.
The quiescent X-ray luminosity $L_{\rm X}$ ($\simeq 5\times 10^{31}$ 
erg s$^{-1}$; \citealt{cam+02,hei+07}) is two orders of magnitude 
lower than that required \citep{bur+03}. Instead, if one considers that 
the neutron star in the binary turns on
as a radio pulsar in quiescence, its rotational energy output (so-called
spin-down luminosity) $L_{\rm sd}$ is 
$\simeq 9\times 10^{33}$ erg s$^{-1}$ \citep{har+08}. This energy output
would presumably be in the form of a pulsar wind and sufficient to illuminate 
the inner face of the companion star.
Among over a dozen known APMP binaries \citep{pat10}, two other systems,
IGR~J00291$+$5934 and XTE~J1814$-$338, were also found to have similar 
optical modulations that cannot be explained by X-ray heating 
(\citealt{dav+07, dav+09}; however for the first source, see also 
\citealt{jts08}), suggesting that APMPs in these systems might commonly
switch to be radio pulsars during quiescence.

To fully examine the possibility in \saxs, Deloye et al. (2008; see also 
\citealt{hei+09}) have observed \saxs\ simultaneously at X-ray and optical
energies, and from the observations they have confirmed the inconsistency
between the large-amplitude optical modulation and low X-ray luminosity.
Applying a phase-coherent timing technique, \citet{wan+09} have accurately 
determined the period and phase of the optical modulation and concluded that
the optical periodicity is highly consistent with the X-ray orbital ephemeris
(derived from pulsar timing; \citealt{har+08}).
Their results have excluded other possible origins (such as 
the accretion disk) and established that the modulation does arise from 
the companion star.

Following these studies, it should be interesting to compare the modulations
between outburst and quiescence since presumably the former is caused by
X-ray heating, and the latter by pulsar wind heating. When \saxs\ 
was reported to have its expected outburst \citep{gal08} at the end 
of 2008 September \citep{ms08}, we organized optical observations of 
the source and obtained its light curves during the outburst.
In this paper, we report on the results from the observations.

In addition, \citet{del+08} have shown that by fitting the quiescent optical 
modulation in \saxs\ with an advanced binary light curve model, in which 
lights from an accretion disk and an irradiated donor are considered, 
important constraints on the binary system 
(such as the inclination and neutron star's 
mass) can be derived. 
Their work also concluded that in order to obtain better parameter constraints
for this system, simultaneous multi-band light curves would be needed. 
Using data newly obtained as well as those reported in \citet{del+08} and
\citet{wan+09}, we have conducted
fitting to the modulation. We used the binary light curve code
{\tt Icarus} \citep{bre+12},
that is similar to the ELC program \citep{oh00} used by \citet{del+08}.
We also employed a Markov chain Monte Carlo (MCMC) technique that allows 
for simultaneously fitting the multi-dimensional light curve model to 
the data and identifies the best-fit parameters as well as their confidence 
intervals.
\begin{figure}
\includegraphics[scale=0.48]{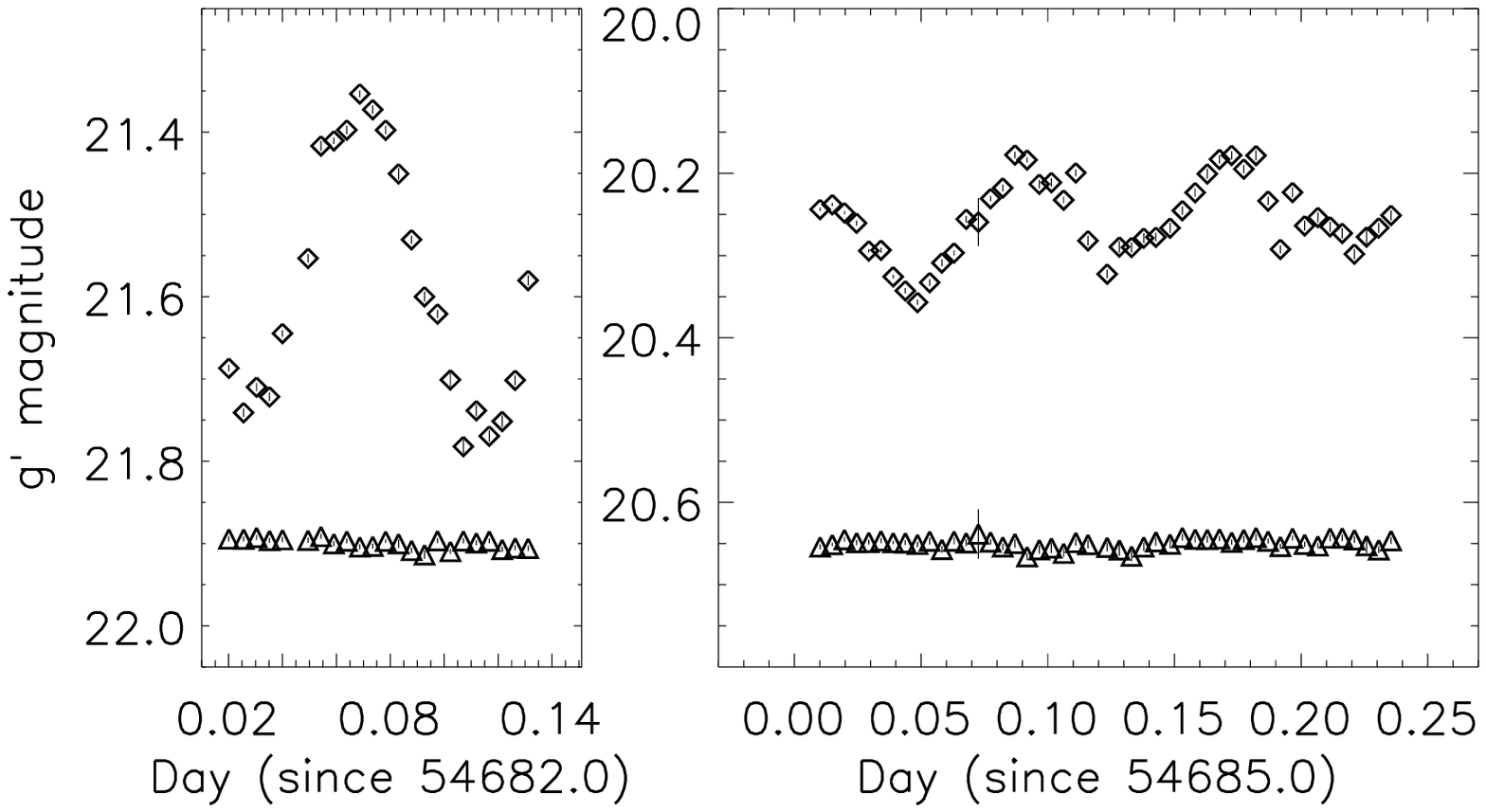}
\includegraphics[scale=0.48]{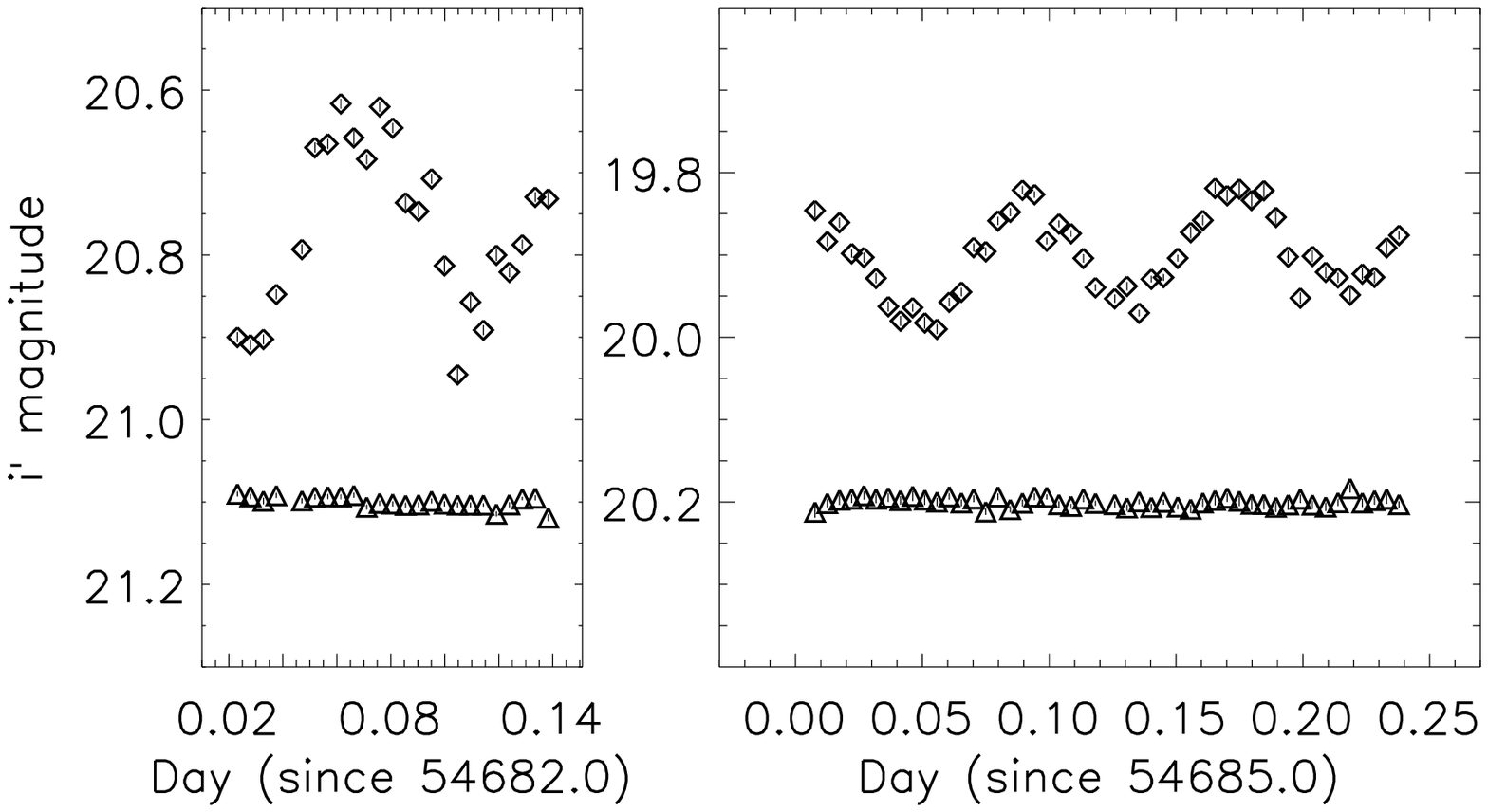}
\figcaption{Gemini South $g'$ (\textit{upper} panels) and 
$i'$ (\textit{bottom} panels) light curves of \saxs\ obtained
in 2008 August. 
Light curves of a comparison star (triangles) are also plotted.
\label{fig:lcgi} }
\end{figure}

We describe the previous light curve data, our time-resolved photometry 
of \saxs\ in 2008, and related data reduction 
in \S~\ref{sec:obs}. We provide details of our binary light curve model 
and MCMC fitting of the obtained light curves, and fitting results 
in \S~\ref{sec:fit}. Timing analysis of all the quiescent light curves and
analysis of broad-band disk spectra obtained from our fitting are given
in \S~\ref{sec:ana}. We discuss interesting results from our observations
in \S~\ref{sec:dis}, and a summary is provided in \S~\ref{sec:sum}.

\section{OBSERVATIONS AND DATA REDUCTION}    
\label{sec:obs}

\subsection{Previous Gemini Data Obtained in Quiescence}
\label{sec:pobs}

Two sets of light curve data obtained from previous Gemini imaging observations
were included in this work. The observations for the first and second set 
were carried out in 2007 March and 2008 May, and were reported
in \citet{del+08} and \citet{wan+09}, respectively. The first set mainly 
contains two $g'$ light curves, one 2.5 hour long and the other 2.8 hour long,
while two additional $i'$ brightness
data points were obtained at the end of each $g'$ light curve. The second set
contains three 3~hour $r'$ light curves obtained over 5 days. 
Detailed descriptions
of the datasets can be found in \citet{del+08} and \citet{wan+09}. 

\subsection{Gemini Imaging in 2008 August in Quiescence}

Time-resolved imaging of \saxs\ was carried out with Gemini South Telescope
on 2008 August 4 and 7. 
To obtain nearly simultaneous multi-band light curves,
Sloan $g'$ and $i'$ filters were alternately used for imaging in each night. 
The detector was the Gemini Multi-Object 
Spectrograph (GMOS; \citealt{hoo+04}), which consists of 
three 2048$\times$4608 EEV CCDs. In our observations, 
only a section of 300$\times$300 
pixel$^2$ in the middle CCD (CCD 02) was used.  The pixel scale 
is 0.073\arcsec/pixel. 

The total observation time in the first night was 3 hours, with
23 and 24 useful $g'$ and $i'$ images respectively
made. The exposure times were 200.5~s in $g'$ and 150.5~s in $i'$.
The observing conditions were variable, with the seeing
[full-width half-maximum (FWHM) of the point spread function (PSF) of 
the images] increasing from 0.6\arcsec\ to 0.9\arcsec\ over the course
of the observation.
On August 7, the total observation time was 5.5 hours, during which 47 and 
48 useful $g'$ and $i'$ images, respectively, were made.
The exposure time at each band was the same as that in the first night.
The observing conditions were relatively stable, with the average seeing
being $\simeq$0.7\arcsec.

\subsection{Canada France Hawaii Telescope (CFHT) Imaging in the 2008 Outburst}

The starting of the 2008 outburst of \saxs\ was reported by \citet{ms08}
on 2008 September 22. We subsequently requested CFHT Target-of-Opportunity 
observations of the source. The observations were carried out in the
queued service observing mode on 2008 September 29, 30, October 3, and 4.
The detector was the wide-field imager MegaCam, which consists of 36
2048$\times$4612 pixel$^2$ CCDs. The field of our target was imaged on 
CCD 22. The pixel scale is 0.187\arcsec/pixel.  A Sloan $r'$ filter was used 
for imaging.

During the four nights, we obtained 60, 63, 84, and 72 images, respectively.
The exposure time for each image was 20~s. The total observation times were
1.1, 1.2, 1.6, and 1.4 hours, respectively. The observing conditions in
the first night were good, with an average seeing of 0.8\arcsec.
In the second night, the seeing condition was not good,
having a large range of 0.8--1.8\arcsec\ and an 
average of 1.1\arcsec. Therefore
6 images with large seeing values were excluded from the data.
The third night had good observing conditions, with an average seeing of
0.7\arcsec. The observing conditions in the fourth night were relatively 
variable, with the seeing in a range of 0.6--1.2\arcsec and an average 
of 0.8\arcsec.

\subsection{Data Reduction and Photometry}

We used the IRAF packages for data reduction. The images were bias subtracted
and flat fielded. 
We performed PSF-fitting photometry to measure the brightnesses of 
the source and other in-field stars.  A photometry program 
{\tt DOPHOT} \citep{sms93} was used.  
\begin{figure*}
\begin{center}
\includegraphics[scale=0.55]{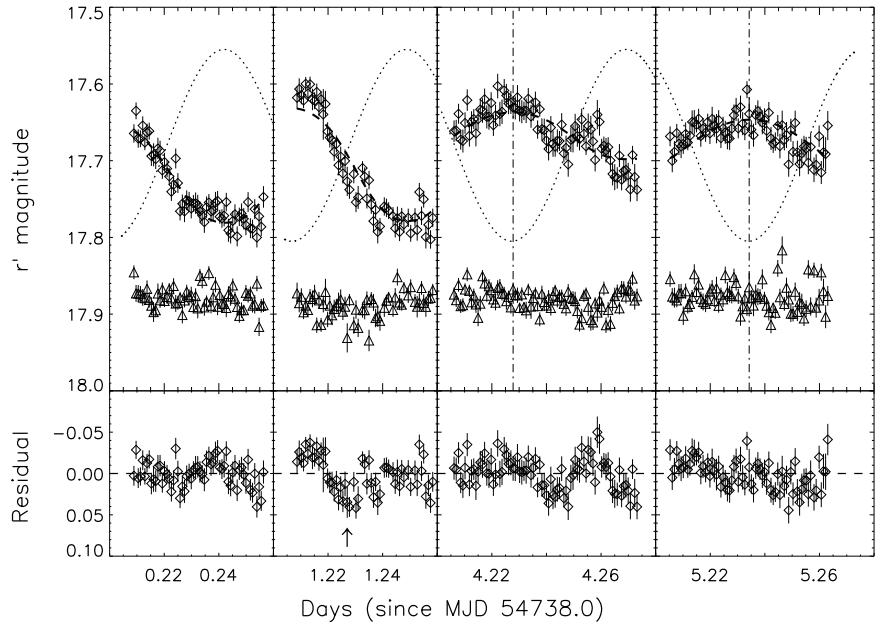}
\figcaption{CFHT $r'$ light curves of \saxs\ in the 2008 outburst (diamonds),
with the last two up-shifted by 0.3 mag. 
For comparison, light curves of a comparison star are plotted as triangles. 
The X-ray ephemeris of the pulsar, which gives its mean orbital longitude
\citep{har+08}, is shown as dotted curves, and positions
of superior conjunction of the companion are indicated in the last two panels
(dash dot lines).
Our model light curves are shown as dashed curves.
In the residuals of the second light curve from our model fit, significant
deviation of a few data points from our model is seen (marked by an arrow),
indicating the modulation peak is narrower than that of our model.
\label{fig:lcr} }
\end{center}
\end{figure*}

Before photometry, we positionally calibrated our Gemini images 
(made during the quiescent state) to reduce possible contamination 
from the two nearby stars,
since the two stars had similar brightnesses and were 0.6\arcsec\ 
and 1.0\arcsec\ away from the target (e.g., \citealt{wan+09}). 
We first made a reference image by combining three best-quality $i'$ images on
August 7. All Gemini images were then calibrated to this reference image. 
We determined the positions of our target and the two nearby stars in the 
reference image and fixed them at these positions for photometry of the Gemini
images. Differential photometry was performed to eliminate systematic flux 
variations in the images. Six isolated, non-variable bright stars in the field 
were used. The brightnesses of our targets and other stars in each frame were
calculated relative to the total counts of the six stars. 
Standard stars 95-100 and 95-96 \citep{lan92} were observed on 
August 4 and used for $g'$ and $i'$ flux calibration, respectively.
To convert \citet{lan92} Vega magnitudes of the standard stars to Sloan filter
magnitudes, transformation equations given by \citet{fuk+96} were used.

For CFHT data, because \saxs\ was in outburst, approximately 30 times
brighter than in quiescence, and the exposure time was short, no effort was 
made to separate the two nearby stars from our target. Differential photometry
was also performed, with 10 in-field bright stars used for calibrating out
systematic flux variations among the images. 
No standard stars were requested in our CFHT program. Using 8 in-field 
bright stars, we flux calibrated our $r'$ images to those in \citet{wan+09}.

\subsection{Light Curve Results}

The obtained Gemini and CFHT light curves are shown in Figure~\ref{fig:lcgi} 
and Figure~\ref{fig:lcr}, respectively. We note that during our Gemini 
observations, the source was brightening. Over two days, the source's $g'$ and
$i'$ brightnesses changed by 1.3 and 0.9 mag, respectively. In addition, the
amplitudes of modulation decreased from $\simeq$0.4 mag to $\simeq$0.2 mag.

At the time of the outburst, \saxs\ was visible to CFHT for only approximately
one hour. Therefore each CFHT light curve covered half of the binary orbit. It
can be seen that the first two light curves are sinusoidal-like with an 
amplitude of 0.2 mag, similar to those seen in quiescence. The latter two 
light curves changed to a smaller amplitude of 0.1 mag, and 
the average source brightness at the time decreased by approximately 0.3 mag.

\section{Fitting}
\label{sec:fit}

\subsection{Light curve model}
We fitted the data using the {\tt Icarus}\footnote{Freely available at
https://github.com/bretonr/Icarus.} light curve model for irradiated 
companions in binaries \citep{bre+12}. The model was adapted from the 
code developed by 
van~Kerkwijk \citep[e.g.][]{sta+01}, and has been recently used for two 
other systems \citep{vbk10,vrb+10,bre+12}. The model is similar to the ELC 
program \citep{oh00} that was employed by \citet{del+08} for their earlier work 
on \saxs.

In this model, the surface of the companion is constructed by solving 
the hydrostatic equilibrium equation for a rotating body, which in our case 
was tidally locked to the neutron star. We implemented a new 
stellar grid parametrization that uses a triangle tessellation obtained 
from the subdivision of the primitives of an isocahedron. This parametrization
yields uniform coverage on the projected sphere surrounding the star and 
allows for better performance of the code. We scaled the star so that its 
size matches that of the Roche lobe, which is justified from the fact 
that \saxs\ regularly undergoes outbursts involving mass transfer from 
the companion. A base temperature for the companion is chosen and effects 
of gravity darkening accounted using Lucy's law \citep{luc67}, with a 
power-law index of 0.08. To this, we apply the irradiative flux from 
the neutron star in the way prescribed by \citet{oh00}. Finally, 
we integrated over the visible surface the emerging flux in a selected 
photometric filter, which was interpolated from 
BTSettl atmosphere models \citep{all+03,all+07,all+11}, in order to obtain 
the total flux at a given 
orbital phase and for specific orbital parameters.
\begin{figure*}
\begin{center}
\includegraphics[scale=0.55]{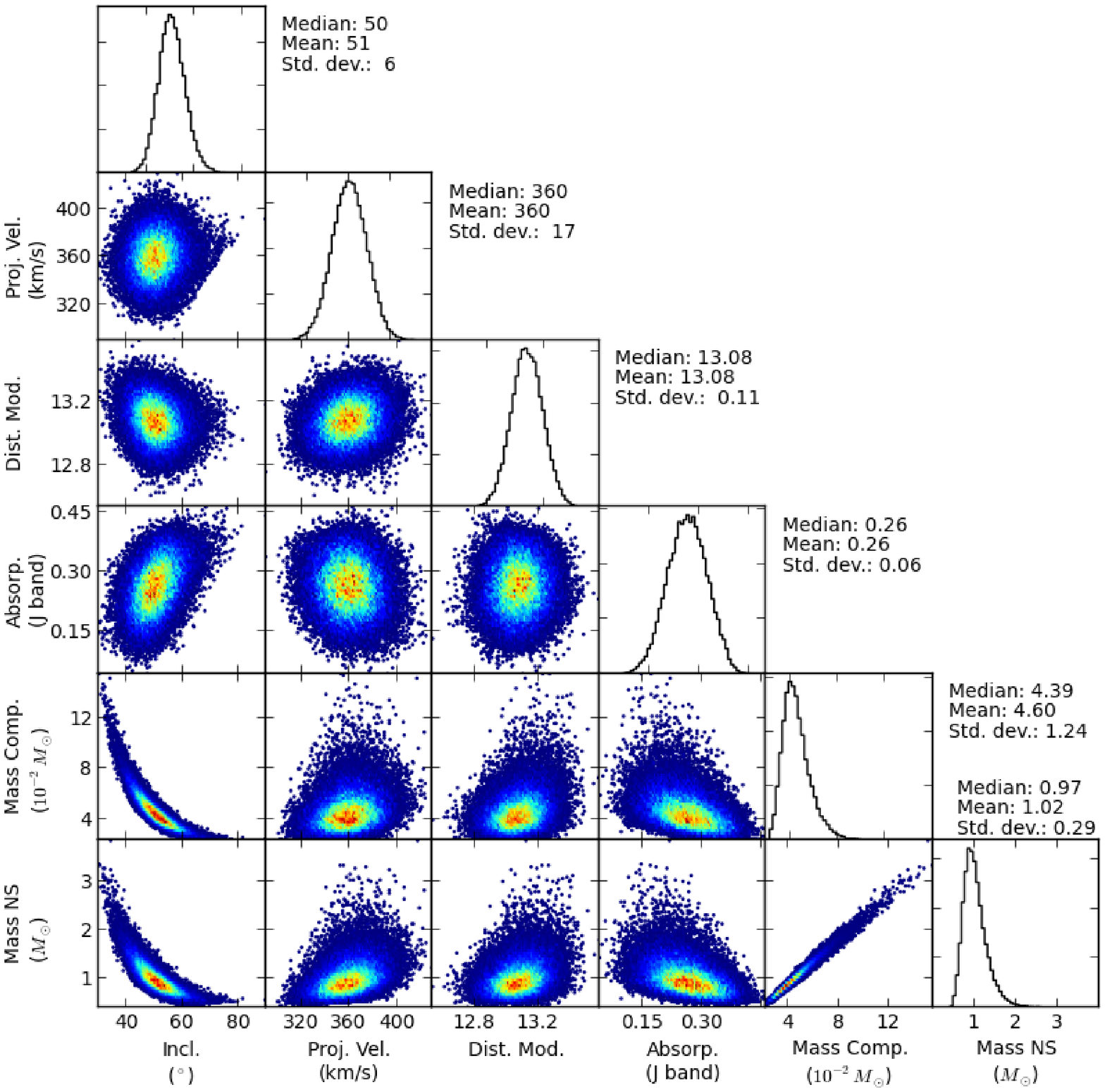}
\figcaption{One- and two-dimensional distributions of the key parameter values
resulting from our MCMC fitting. Values of the median, mean, and standard 
deviation of each parameter are given. The masses are inferred values
from the neutron star mass function, the orbital inclination, and the projected
velocity.
\label{fig:mcmc}}
\end{center}
\end{figure*}

We approximated the disk contribution by adding an orbital-independent, constant
value to the companion's flux. 
The disk flux values inferred at different photometric bands at one 
epoch constitute
a spectral energy distribution, which were checked
for consistency with arising from an accretion disk (\S~\ref{subsec:disk}).
We also made the implicit presumption that 
the system is seen sufficiently face-on so that there is no mutual 
eclipse/shadow arising between the disk and the companion. This presumption
is motivated by the fact that light curves of \saxs\ in quiescence are 
symmetrical and do not show obvious signs of flux changes that could result 
from the interaction between the two components \citep{wan+09}. 
Moreover, the estimated 
orbital inclination from previous work (e.g., \citealt{del+08}) as well as 
that inferred from this work (see \S~\ref{subsec:results}---the maximum value
we found is only 56\arcdeg) indicates that 
the companion is unlikely to be obscured by the disk for any reasonable 
disk scale height.  

\subsection{Bayesian inference}

We used a Bayesian framework to make statistical inferences on our light 
curve modeling parameters. The Bayesian inference on a set of parameters 
$\vec{\theta}$ is obtained from their posterior probability, which is 
conditional to the prior knowledge, $I$, that we have of these parameters 
and the experimental data $D$. The posterior probability can be written as 
the product of the priors times the likelihood of the data: $p(\vec{\theta} | I, D) = p(\vec{\theta} | I) \,\, p(D | \vec{\theta}, I) / p(D|I)$, 
where the denominator is a normalization factor to ensure that the posterior 
probability integrated over the parameter space is unity [see \citet{gre05} 
for an introduction to Bayesian analysis].

\subsection{Model parameters}
\subsubsection*{Inclination and Radial velocity semi-amplitude}\label{sec:RV}

The X-ray timing of \saxs\ already provides accurate measurements of its
orbital period and the 
projected semi-major axis of the neutron star. In order to
fully parametrize the system (e.g. masses, separation, etc.), two additional 
quantities are required. We chose the orbital 
inclination, $i$, and the projected radial velocity semi-amplitude of 
the companion $K_{\rm comp}$, because of their close connection to observables.
For the orbital inclination, \citet{del+08} found a relatively large range of
36--67\arcdeg\ (1$\sigma$) from their fitting. Since a subset of the data
used in our analysis included the data set from Deloye et al. (2008),
we chose the priors to be flat in $\cos i$ to fully search
for the best-fit range in this work.

The use of spectroscopic observations to constrain the companion's radial 
velocity is made quite difficult due to its faintness in quiescence. 
While the system does get much brighter during outbursts, 
its emission mainly contains features arising from the accretion disk. 
Using spectra obtained in the 2008 outburst and analyzing Doppler 
images of the \ion{N}{3} $\lambda 4640$ emission line associated 
to the Bowen blend, \citet{cor+09} and \citet{ele+09}
both tracked the motion of the companion in \saxs.
Because the latter analyzed a total of 39 spectra, which included 16 spectra
that were analyzed and reported by the former,   
we considered the result given by the latter, 
$K_{\rm Bowen} = 324 \pm 15$\,km\,s$^{-1}$, and reported our fitting
results based on this measurement. We discussed the influence of 
the different $K_{\rm Bowen}$ measurements to our fitting in 
\S~\ref{subsec:d1} by also considering the value obtained 
by \citet{cor+09}, $K_{\rm Bowen} = 248 \pm 20$\,km\,s$^{-1}$. 
However only unreasonably low-mass values for the neutron star
were found from our fitting when the value of \citet{cor+09} was used.

The average location of the Bowen emission is indicated by $K_{\rm Bowen}$,
which is essentially produced on the irradiated hemisphere of the star and 
has a different velocity from that of the center of mass, $K_{\rm comp}$. 
\citet{ele+09} discussed the ``K-correction'' factor required to convert 
the Bowen velocity to the center of mass and estimated that 
$K_{\rm comp} = 370 \pm 40$\,km\,s$^{-1}$, based on modeling 
by \citet{mcm05}.

We included the $K_{\rm Bowen}$ constraint as a Gaussian prior in our 
Bayesian inference. For every realization of the model, we determined the 
velocity of the companion's light center by calculating the average velocity 
over the visible surface, weighted by the flux contribution of each surface 
element. Since the irradiated side of the companion is much hotter than that 
of the back side (6000\,K vs. 3000\,K, respectively), the flux contribution 
from the non-irradiated side in the $g'$-band becomes negligible and 
the inferred velocity should be a good approximation of $K_{\rm Bowen}$. 
As expected from \citet{mcm05}, we found that 
$K_{\rm comp}\sim 370$\,km\,s$^{-1}$ is required to satisfy the velocity 
constraint. Such value is also consistent with the semi-analytical 
K-correction relationship derived by \citet{vbk10} for the companion 
to PSR~B1957+20, which shares similarities to \saxs. From analytical and 
numerical modeling of this irradiated companion, they determined that 
the ratio of the observed projected light-center velocity to the 
projected center-of-mass velocity 
follows $K_{\rm obs} / K_2 = 1 - f_{\rm eff} R_{\rm nose}/a_2$,
where $R_{\rm nose}/a_2$ is the radius of the nose of the companion in units 
of separation to the system's barycenter, and $f_{\rm eff} \sim 0.6$, a 
parameter relatively independent of the system and companion properties 
such of orbital inclination, filling factor, and surface temperature 
as long as the companion is strongly irradiated and nearly fills its Roche 
lobe. Hence, for \saxs\ $R_{\rm nose}/a_2 \sim 0.2$ which, again, implies 
a measurable velocity comparable to the above value.

\subsubsection*{Base and day-side temperatures}

The base temperature of the companion was fixed using a temperature-mass 
relationship for low-mass brown dwarfs \citep{del+08}.
The temperature, as a function of mass alone, ranges between 2000--3000 K.
From the photometric colors near the inferior conjunction of 
the companion, we confirmed that the temperature is consistent with the range.
In addition, keeping it a free parameter did 
not improve the light curve fits significantly.

We used flat priors for the companion's day-side temperature since its 
surface albedo is uncertain as well as the amount of incident energy 
approximated 
by the pulsar's spin-down energy. Although it is unlikely, we tested 
modeling the whole dataset with a constant day-side temperature. We found such
a setting did not result in good fits. 
A separate day-side temperature for each observation was used in order to 
account for variability.

\subsubsection*{Distance modulus and reddening}

Using several different methods, \citet{gc06} estimated the distance $d$
to \saxs. First by setting the time-averaged X-ray flux equal to the
expected mass transfer rate from gravitational radiation, they found a
low limit of $d>3.4$~kpc.
They also modelled the type I X-ray bursts that all showed 
photospheric radius expansion during the 2002 October outburst of \saxs,
and obtained a distance range of $d = 3.1-3.8$\,kpc.
Finally considering the peak flux of the radius expansion bursts
as a standard candle, $d=3.6$~kpc was estimated for a pure helium atmosphere
or $d=4.3$~kpc from an empirical value of \citet{kuu+03}.
We set a Gaussian prior on the distance 
modulus ${\rm DM} = 12.72 \pm 0.15$, where a  
7\% uncertainty was chosen in order to compensate for possible 
uncertainties in the distance estimates (see \citealt{gc06} for details).

For the reddening, we also chose a Gaussian prior and set the $J$ band 
absorption to $A_J = 0.21 \pm 0.05$ using the value derived from \citet{wan+09}.

\subsubsection*{Disk flux}

The same as for the day-side temperature, we used an independent disk 
flux value with a flat prior for each set of continuous observation, as 
well as each photometric band. We designed our light curve fitting algorithm 
to optimize the disk flux value every time the companion parameters were 
changed. Since the disk contribution is an additive, linear problem in 
the flux system --- and therefore a very simple non-linear problem in 
the magnitude system --- solving for the best disk flux comes to 
the cost of only a few additional CPU cycles compared to the lengthier 
optimization of the companion parameters.
\begin{figure*}
\begin{center}
\includegraphics[scale=0.7]{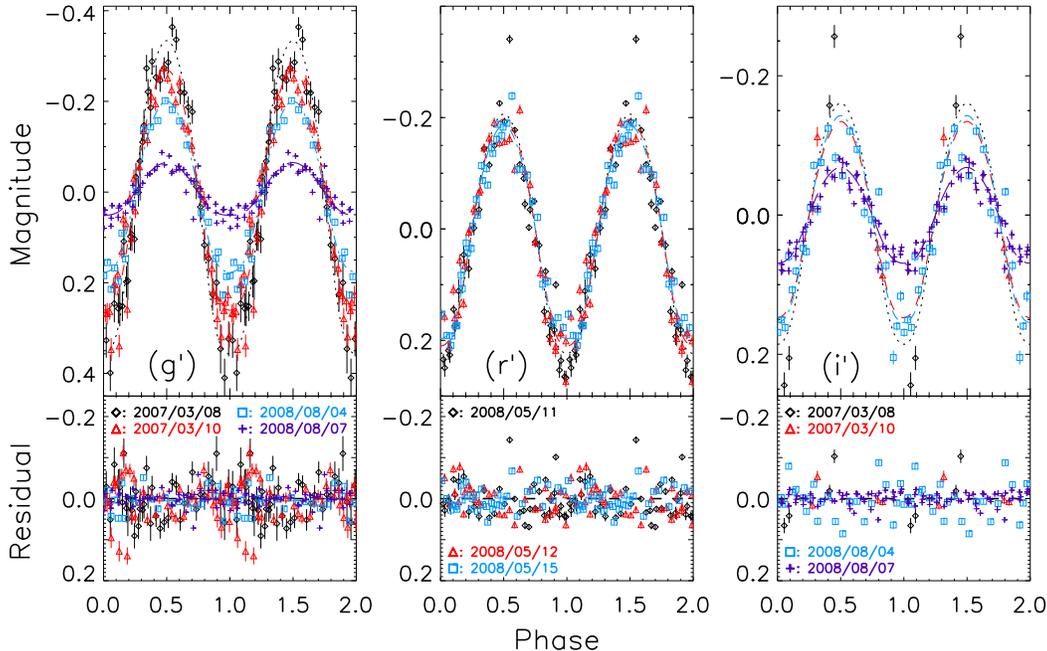}
\figcaption{Folded quiescent light curves (at $P=7249.157$~s), 
with $g'$, $r'$, and $i'$
data shown in the left, middle, and right panel, respectively. Following
the data sequence given in Table~\ref{tab:res}, the data are plotted as
diamonds, triangles, squares, and plus signs, respectively, while the best-fit
model light curves are plotted as dotted, dashed, dash-dot, and long dash
curves, respectively. Two cycles are displayed for clarity.
\label{fig:fits}}
\end{center}
\end{figure*}

In the 2008 August 7 observation (Figure~\ref{fig:lcgi}), a noticeable 
rise in the overall flux 
level is visible and likely indicates that \saxs\ was coming out of 
quiescence. Clearly, a constant flux value cannot account for this change 
and we allowed for a small linear variation of the disk contribution over 
the course of the observation.

\subsubsection*{Noise parameter}
\label{sec:noise}
In order to account for intrinsic short-term flux variations that sometimes
could significantly deviate away from the sinusoidal-like orbital modulation 
of \saxs\  (\citealt{del+08,wan+09}; see also \S~\ref{subsec:orb}), 
we added a `noise parameter' to the uncertainties of our data.
The uncertainty of a flux datum $i$ in dataset $j$ becomes 
$s_{ij} = \sqrt{b_j} \sigma_{ij}$, 
where $s_{ij}$ and $\sigma_{ij}$ are the new and old uncertainties, 
respectively, and $b_j$ is the noise parameter. This technique is commonly 
used in Bayesian analysis (see, e.g., \citealt{gpt99,gre11}) and has an 
effect loosly similar to that of uncertainties rescaling which is performed 
in frequentist analysis to obtain a reduced $\chi^2 \sim 1$. 
The priors are chosen to be logarithmic (i.e. Jeffrey's priors) in order 
to provide equal weighting per decade since the noise term is a scaling 
parameter. 

\subsection{MCMC fitting}

The light curve fitting is a multi-dimensional nonlinear optimization problem 
that we tackled using an MCMC 
method \citep{grs95}. This is particularly efficient to make Bayesian 
inference since the full posterior distribution is evaluated. In the limit 
that the observed magnitude errors are normally distributed, the likelihood 
function can be written as $p(D|\vec{\theta}I) \propto \exp (-\chi^2/2)$, 
where $\chi^2$ is the conventional chi-square.

For the MCMC, we chose the \emph{stretch move} algorithm that is part of a 
family of MCMC methods called \emph{ensemble samplers with affine invariance} 
described by \citet{goowea09a}. In a nutshell, the algorithm is executed by 
simultaneously running several chains, all initialized at random locations 
of the parameter phase space. Every step of the MCMC, a move is proposed for 
each chain by choosing a complementary chain at random and drawing point 
along a line passing through the last position recorded in the two chains. 
How big a step away from the previous position is determined using a 
distribution which is affine invariant [i.e., $g(z^{-1}) = z g(z)$]. 
The proposed move is accepted with a probability that is 
slightly different than that of the usual Metropolis 
algorithm in order to satisfy the detailed balance requirement 
[see \citet{goowea09a} for the details]. An advantage of the stretch 
move algorithm is that the acceptance rate, 
which controls the efficiency of the chain, can be tuned using a single 
parameter in the proposal distribution, as opposed to one per dimension 
for a random walk Metropolis algorithm. It should also perform better 
for highly skewed and badly scaled distributions.

We ran the stretch move algorithm using 30 simultaneous chains for a total 
of $100\,000$ steps each. The proposal distribution's tuning parameter was 
determined from a previous trial run and the obtained acceptance rates of 
the chains were found to be in the range $20-30\%$, as prescribed 
by \citet{rr01}. For subsequent analysis, we discarded the 
first $30\,000$ points (e.g., the burn-in period) and thinned the remaining by 
keeping every other 50 points in order to reduce the auto-correlation. 
In order to ensure that our MCMC has converged we visually inspected the trace
of the parameters and performed the Geweke 
test \citep{gew92} by comparing the mean of 10\%-subsections of the first 
half of the chain to the remaining second half.

\subsection{Results}\label{subsec:results}

We originally aimed to compare the modulations
in quiescence and outburst by fitting the light curves separately.  
However the combination of the short light curves during the outburst 
and their relatively large uncertainties does not provide tight constraints.
We also tested fitting all data together, but for the same reason
the results were hardly changed compared to those including the
quiescent data only.
We therefore fit all the quiescent data obtained from 2007 March to 2008 
August with our model.
In Table~\ref{tab:res}, we provide a summary of the fit results as well
as several additional quantities inferred from the marginalized posterior
probabilities of our irradiated companion modeling.
The one and two-dimensional posteriors are displayed in Figure \ref{fig:mcmc}.

We show all folded light curve data points, model light curves of 
the median parameter values, and residuals in Figure~\ref{fig:fits}. 
As can be seen, our model generally reproduces the observations quite 
well with
the data points evenly distributed along the model light curves.
However it can be noted that a few light curves have relatively large
variations around our model fits, which are also reflected by the noise
parameter values.
These variations, noted in previous studies \citep{del+08, wan+09}, 
were likely intrinsic and stochastic and can not be
described by our current model.
\begin{center}
\includegraphics[scale=0.6]{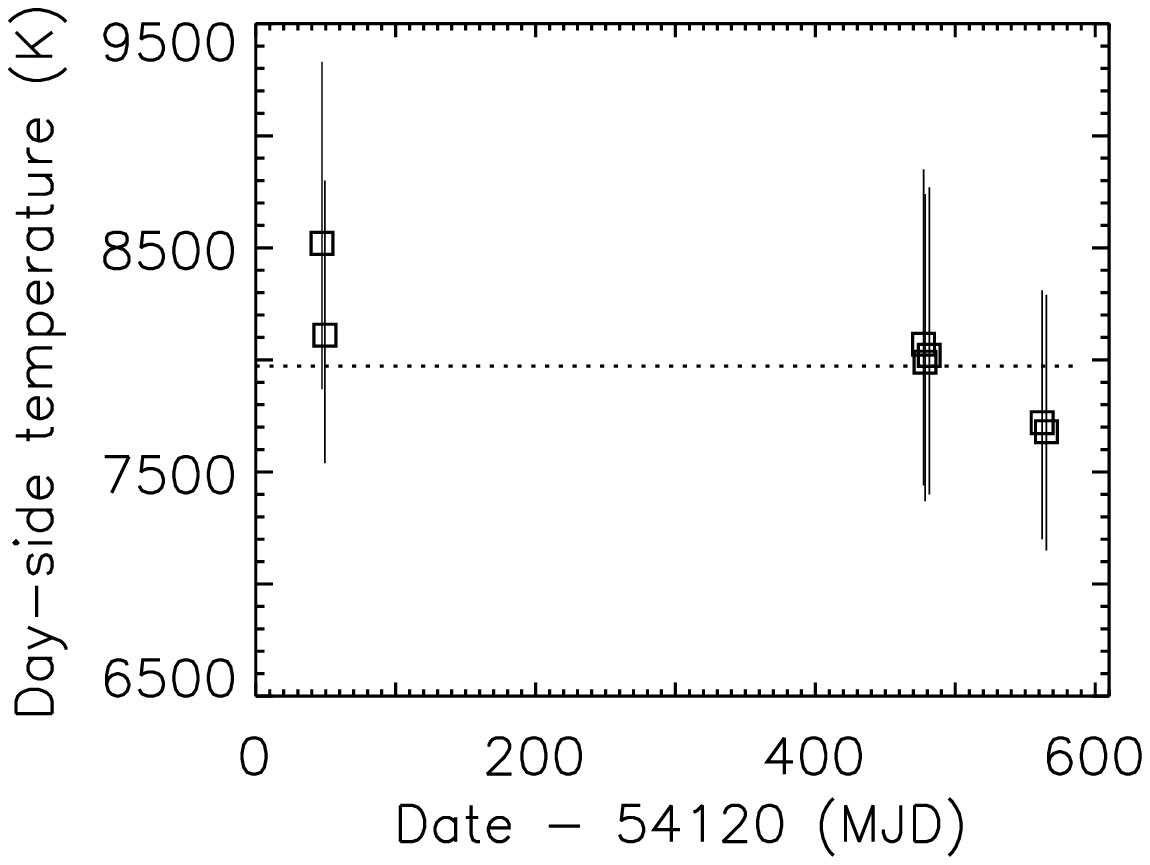}
\figcaption{Day-side temperature of the companion star in \saxs\  as 
a function of time, which can be described by a constant of 7970~K 
(dotted line).
\label{fig:temp} }
\end{center}

Our inferred orbital inclination, $i = 50^{+6}_{-5}$ deg, 
is consistent 
with previous estimates from \citet{del+08} and
\citet{cac+09} (the latter found $i=55^{+8}_{-4}$ deg by 
relativistic Fe line fitting; see also \citealt{pap+09,pat+09})
despite the day-to-day disk and irradiated-side temperature variability, 
which probably impair our ability to combine multi-epoch data obtained 
in different bands. Consequently we constrained the neutron star and
companion to have masses of 0.97$\pm^{+0.31}_{-0.22}\ M_{\sun}$ and 
0.04$^{+0.02}_{-0.01}\ M_{\sun}$, respectively. 
Our fitting has narrowed the possible mass ranges for both the neutron star 
and companion and points to a low-mass neutron star in \saxs, while 
we note with caution that the 3$\sigma$ range for the mass of the
neutron star is still wide, 0.47--2.52 $M_{\sun}$ (Table~\ref{tab:res}).
In addition, we note that our measurement of $M_{\rm ns}$ is 
in agreement with the independent estimate of $M_{\rm ns}<1.5\ M_{\sun}$ from 
X-ray pulse shape modeling (\citealt{ml11}).

The results of the distance modulus and extinction,
DM=13.08$\pm$0.11 and $A_J=0.26\pm0.06$,
are slightly larger than the previous estimates. 
The posteriors are very close to being normally distributed, and 
have standard deviations nearly the same as that of the priors.
Our inferred companion's projected radial velocity 
semi-amplitude, $360^{+17}_{-16}$\,km\,s$^{-1}$, is in agreement with 
the value, $370 \pm 40$\,km\,s$^{-1}$, from \citet{ele+09}. We observed 
that the uncertainty on $K_{\rm comp}$ is only 50\% as large as that of 
the priors, which implies that the data contribute 
to constraining the companion's velocity. The most likely reason is 
that given the distance and absorption are relatively well-determined, and 
the fact that the companion fills its Roche lobe, the observed luminosity 
constrains the actual radius of the companion which is tied up to its 
orbital velocity.
\begin{center}
\includegraphics[scale=0.6]{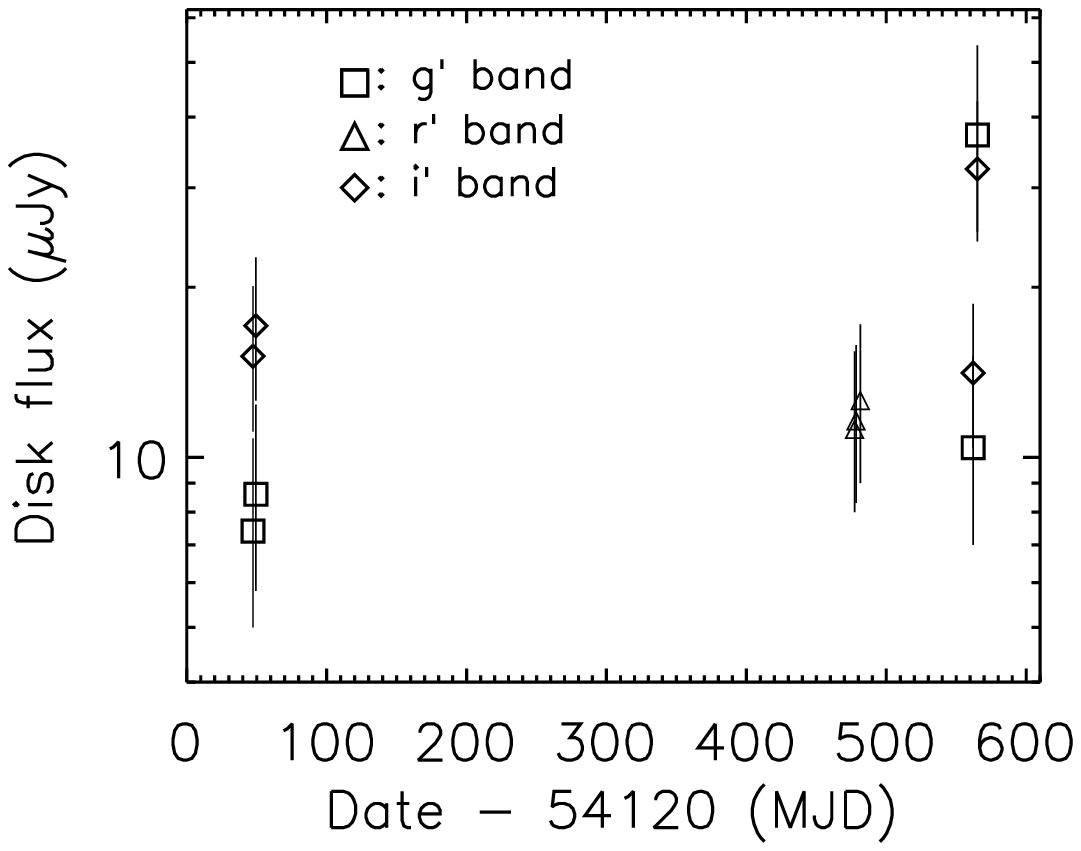}
\figcaption{Disk flux contribution as a function of time. The squares, 
triangles,and diamonds indicate the $g'$, $r'$ and $i'$-band data, respectively.
\label{fig:disk} }
\end{center}

The day-side temperatures of the companion 
are consistent with being a constant before the 2008 outburst
(Figure \ref{fig:temp}),
and we found that a temperature of 7972$\pm$243~K best fits them
($\chi^2= 1.1$ for 6 degrees of freedom).
We note that the data points may suggest a trend of temperature decreasing
over the time, but no conclusion can be drawn due to the large standard 
deviations.  Our result is relatively surprising as 
one might have naively expected increased activity as the system was going 
towards an outburst. 
The behavior of the disk flux contribution (Figure \ref{fig:disk}) is 
also consistent with being constant in each band over time, except 
for the last observation (on 2008 Aug 7) in which the flux increased 
by a factor $\gtrsim 2$ within 
3 days. Detailed analysis of the disk components is given below in 
\S~\ref{subsec:disk}.

Previously by fitting the first two sets of data (obtained on 2007 March 8
and 10; Table~1), \citet{del+08} found $M_{\rm ns}>1.8\ M_{\sun}$, which
favoured a high-mass neutron star in \saxs.
Our results are drastically different, now pointing to a light,
$M_{\rm ns}<1.3\ M_{\sun}$ neutron star. In addition to the constraint
on the companion's radial velocity from the $K_{\rm Bowen}$ measurements
(see also the discussion in \S~\ref{subsec:d1}),
the causes of the difference can be understood from the following.
Nearly simultaneous multi-band light curves are crucial in order to break 
parameter degeneracies and, for instance, tell the stellar temperature 
apart from the orbital inclination \citep{bre+12}, while
\citet{del+08} had very limited multi-band information of the binary 
(see \S~\ref{sec:pobs}).
The other important aspect is that the disk contribution acts as a DC 
component on an observed light curve and is almost completely degenerate 
with the orbital inclination. However, we were able to disentangle the orbital
inclination from the disk in our data since the contribution from the latter
varies over time whereas the orbital inclination remains constant.
Finally our MCMC fitting allowed to efficiently explore the entire parameter 
space whereas 
\citet{del+08} restricted their fitting to well-defined regions with
a coarse grid.

\subsubsection{Outburst results}

Because of the reasons given above, 
we studied the outburst light curves by fitting them with all parameters 
of the binary system fixed at the median parameter 
values obtained from our fitting to the quiescent data (Table~\ref{tab:res}). 
The free parameters were only the day-side temperature 
of the companion and disk flux.
In addition, to have sufficient constraints, the last two sets of the data
were assumed to have one day-side temperature value (the average magnitudes
of the source in the two days do not have a significant change). 
Our model generally fits
the observed light curves, as can be seen in Figure~\ref{fig:lcr}, which 
shows the model light curves as well as residuals of the observed 
data points from them.
From fitting, we found that the day-side temperature was in a range 
of 12,000--25,000 K
and disk flux approximately 20--30 times larger than those in quiescence.
These results are also given in Table~\ref{tab:res}.
A test to use the fitting results from the quiescent data as priors
was conducted, and consistent results were obtained.
\begin{center}
\includegraphics[scale=0.65]{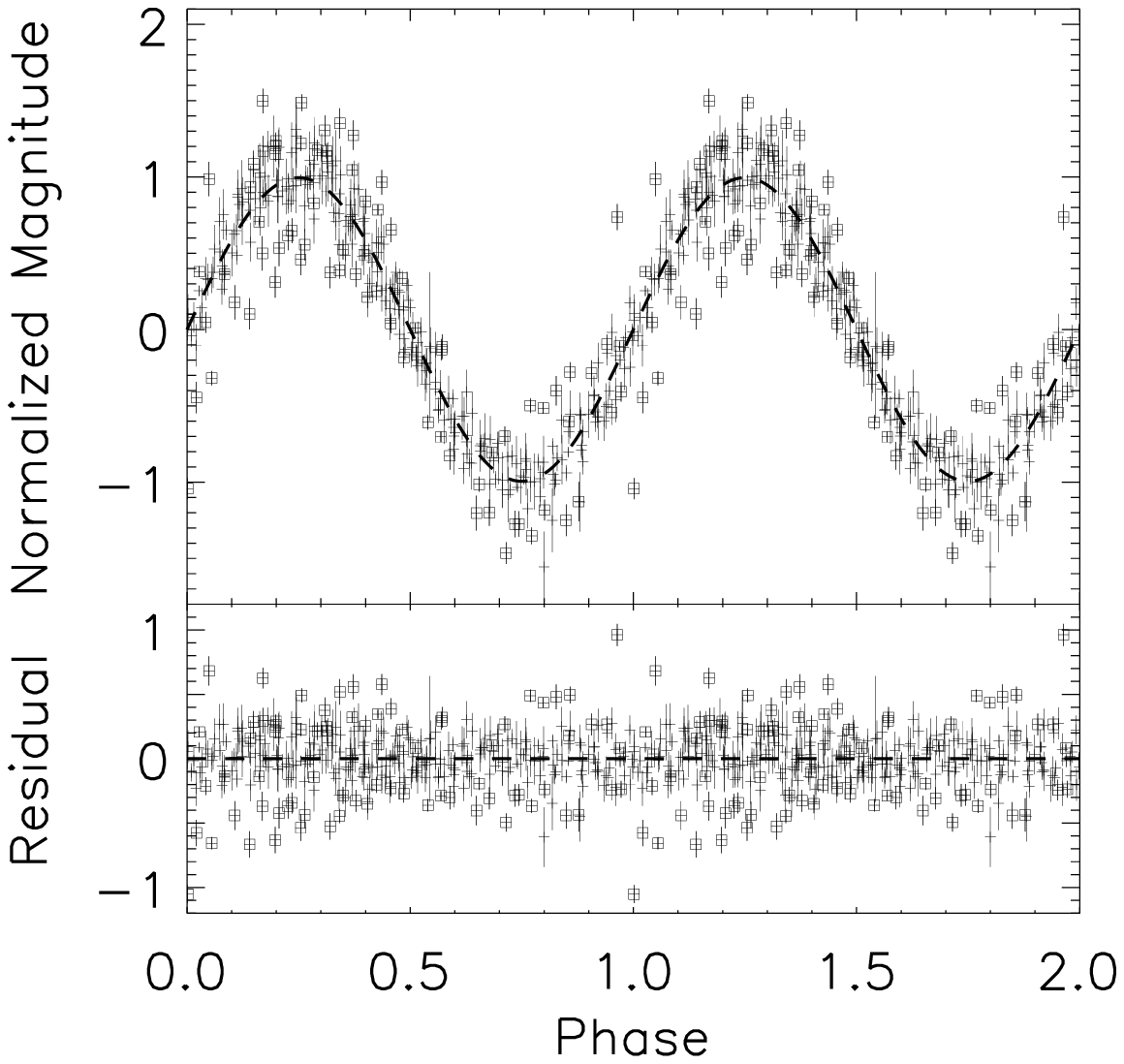}
\figcaption{Normalized and folded multi-band light curves of \saxs. 
The source's multi-band 
light curves can generally be described by a sinusoid function, but contain
a significant fraction of outliers. The data points marked by squares are
those deviate from the best-fit sinusoid more than 3$\sigma$ away.
\label{fig:sin}}
\end{center}

We note that while the overall shape of the modulations in the outburst 
is in general consistent with our model of an irradiated companion star,
there are fine structures in the outburst light curves.
As can be seen in Figure~\ref{fig:lcr}, the second
light curve is sinusoidal-like, but it appears to 
deviate systematically in the middle part from our model.
In addition, the data points in the third and fourth ones are not accurately 
represented and the third one shows a second minor peak after the main peak.
We conclude that minor modulation variations were detected in the outburst.

\section{Analysis}
\label{sec:ana}
\subsection{Orbital Period Determination}
\label{subsec:orb}

It has been shown that the orbital periodicity of \saxs\ can be determined 
from time-resolved photometry with a phase-coherent timing technique
\citep{wan+09}. Using three observations made in 5 days, \citet{wan+09} have
found an uncertainty of 2.8~s on the orbital period $P$. With this uncertainty,
it can be inferred that such timing observations within a maximum time span 
of 3.6 month are needed in order to avoid losing track of the optical 
periodicity phase. Because
our Gemini observations in 2008 August satisfies this requirement, which
further improves the accuracy of the period measurement, we attempted
to phase connect all our light curves obtained during the quiescent state. 
The orbital period of \saxs\ has been found to increase at a rate of 
1.2$\times10^{-4}$ s yr$^{-1}$ from pulsar timing 
\citep{har+08,di+08,bur+09,har+09,pat+12},
and this large value, which is not well understood because it is an order of 
magnitude larger than that predicted by 
standard theoretical calculations for such a binary 
(however see \citealt{pat+12}), might be confirmed from 
optical observations.

Since the modulation was variable over different bands and times of 
the observations (Figure~\ref{fig:fits}), we 
fit each set/band of the light curves individually with a sinusoidal 
function $m=m_c + m_h\sin [2\pi (t/P + \phi_0)]$ (where $t$ is time, 
$\phi_0$ the starting phase, and $P$ is the orbital period fixed at 7249.157 s),
subtracted the obtained constant magnitude $m_c$ from them, and normalized 
them with the obtained semi-amplitudes $m_h$. The uncertainties resulting
from this normalization were added in quadrature to the original uncertainties
of the magnitudes. 
The normalized light curves were then fit with a single sinusoid again, and
we found that the best-fit has an orbital period $P=7249.151\pm0.003$~s, while
$\chi^2=3170$ for 331 degrees of freedom. The large $\chi^2$ value reflects
systematic uncertainties resulting from photometry, intrinsic scattering
of the data points from a single sinusoid \citep{wan+09}, and probably errors 
from normalizing short light curves. 
In Figure~\ref{fig:sin}, the folded light curve, best-fit sinusoid, and 
residuals to the sinusoid are shown. A few data points deviating more
than 3$\sigma$ from the sinusoid are marked by squares. 
We estimated the uncertainties by scaling them
by $(\chi^2/{\rm DoF})^{1/2}$
and obtained $P=7249.150\pm0.008$~s 
and $\phi_0=0.675\pm0.003$ at MJD 54599.0 (TDB), where 
phase $\phi =0.0$ corresponds to the ascending node of the pulsar orbit. 
These results are consistent with that obtained from 
pulsar timing \citep{har+09} within the uncertainties. 
Since the measurement accuracy of a period is generally 
proportional to the length of the time span, it is unlikely to be able
to measure the orbit change rate via long-term optical photometry of 
the source.

\subsection{Accretion Disk During Our Observations}
\label{subsec:disk}

The accretion disk components resulting from our MCMC fitting are shown 
as a function of wavelengths  in Figure~\ref{fig:diskspec}.
The data points 
before the onset of the brightening on 2008 Aug. 7 
can be reproduced by a simple accretion disk model, whose structure 
in quiescence is known to be different from that of the
standard, steady-state disk model (e.g., \citealt*{dhl01}; \citealt*{bb04}). 
In both states the temperature profile can be described by an exponential
function, $T(r)\propto r^{\xi}$, where $r$ is disk radius, but the former
is usually flatter than the latter ($\xi=-0.75$ is the standard value 
for the latter case). 
We fit the quiescent data points with the temperature profile. Free
parameters were disk temperature at $r=10^{10}$~cm, $T_{10}$, and $\xi$.
Source distance, reddening, and inclination angle were fixed at the median 
values found from our MCMC fitting, and the inner and outer edges of the disk 
were assumed to be at the light cylinder of 
the pulsar ($\simeq 1.2\times 10^7$ cm) and 90\% of the Roche-lobe
radius of the pulsar ($\simeq 3.2\times 10^{10}$ cm; \citealt{fkr02}), 
respectively. We
found that $T_{10}=6200^{+100}_{-200}$~K and $\xi=-0.5\pm0.1$ provide 
the best-fit (the minimum $\chi^2=0.3$ for 7 degrees of freedom). 
The small $\chi^2$ value
reflects the large standard deviations of the data points. 
The spectrum of the best-fit disk model is shown as the solid curve in 
Figure~\ref{fig:diskspec}.
\begin{center}
\includegraphics[scale=0.7]{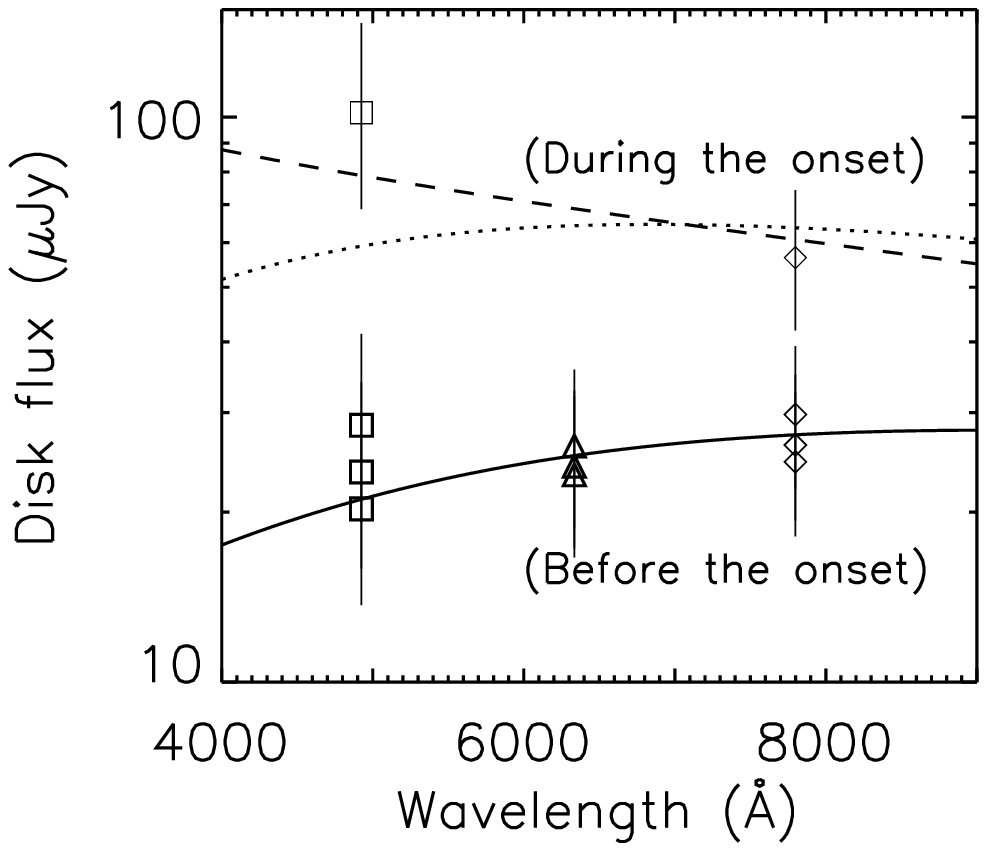}
\figcaption{Broad-band spectra of the disk component in \saxs. Squares, triangles, and
diamonds indicate the $g'$, $r'$, and $i'$ data, respectively. The spectral 
data points
before the onset of the brightening (from 2007 Mar 8 to 2008 Aug 4) can be 
described by an accretion disk that has
temperature profile $T(r)\propto r^{\xi}$ with $\xi=-0.5$ (solid curve). 
During the onset (on 2008 Aug 7), a $\xi=-0.5$
model (dotted curve) can still describe the onset data points, while
a steeper, $\xi\sim-0.8$, temperature 
profile (dashed curve) can also be considered.
\label{fig:diskspec}}
\end{center}

The data points on Aug. 7 are 2--4 times brighter than those obtained
three days earlier, indicating the onset of the disk brightening. In addition,
there is possibly a spectral change from rising to falling
although we have only two data points and their standard deviations are 
too large for such a conclusion to be drawn. For example if
$\xi=-0.5$ is fixed (the same as that obtained above in quiescence),
the disk model with $T_{10}=8200\pm700$~K 
provides a best fit (the minimum $\chi^2= 1.4$ for 1 degree of freedom), 
which is plotted as the dotted curve in Figure~\ref{fig:diskspec}.
For a comparison, a steeper profile such as $\xi=-0.8$ 
($T_{10}=8300$~K) is also plotted (the dashed curve) in the figure.
We note that the brightness at the time 
(average $i'\simeq 19.9$) was approximately 7 times lower than that at the
outburst peak (average $i'\simeq 17.8$; \citealt{ele+09}). 
It is of great interest to study how the disk evolved during the one 
and a half month period from Aug 7 to
September 22 directly preceding the reported X-ray outburst detection. 
Since \saxs\ is known to have an outburst every 3 yrs
\citep{gal08}, a close monitoring of the source before an outburst might
help us learn the disk evolution over such a period.  

\section{Discussion}
\label{sec:dis}
\subsection{Influence of the Companion Radial Velocity}
\label{subsec:d1}

Since the orbital period $P$ and the projected semi-major axis 
of the \saxs\ pulsar 
have been accurately measured from X-ray timing \citep{cm98, har+09}, 
the semi-amplitude 
of the pulsar's projected velocity $K_1$ is known. Combining with it, 
the $K_{\rm comp}$ measurement from optical spectroscopy provides 
a strong constraint on properties of the binary, particularly the neutron 
star's mass $M_{\rm ns}$. It can be shown that 
\[
M_{\rm ns}=\frac{1}{2\pi G}\frac{K_{\rm comp}^3P}{(1+K_1/K_{\rm comp})^2}
	\frac{1}{\sin^3i}\ \ \ ,
\]
where $G$ is the gravitational constant. If $K_{\rm comp}$ is known, 
$M_{\rm ns}$ is determined only by the orbital inclination $i$ 
(similar discussions
were also given by \citealt{cor+09} and \citealt{ele+09}). 
This $M_{\rm ns}$--$i$ relation is 
shown in Figure~\ref{fig:mns}, with 68\% and 99.7\% confidence-level regions
resulting from our fitting (low left panel in Figure~\ref{fig:mcmc}) 
over-plotted. As can be seen, our fitting results
are consistent with the analytical expectations. 

If $K_{\rm comp}= 299\pm$23 km~s$^{-1}$ instead, as reported by
\citet{cor+09}, the $M_{\rm ns}$--$i$ relation is expected to shift
toward lower values. 
We explored this effect by fitting our model using 
$K_{\rm Bowen}=248\pm 20$ km~s$^{-1}$ (\citealt{cor+09})
as a prior. As expected, we found similar results but noticeably 
$i=45\pm5\arcdeg$ and $M_{\rm ns}=0.58^{+0.20}_{-0.14}$\ $M_{\sun}$ 
(both uncertainties are 1$\sigma$), which are both smaller than the values
obtained using the Elebert et al. (2009) measurement as the prior. 
The neutron star now has an unrealistically low mass.
In Figure~\ref{fig:mns}, we show the obtained $M_{\rm ns}$--$i$ regions
from our fitting in detail.
As a result of the change of $K_{\rm Bowen}$ to the smaller value, 
the regions are shifted accordingly, but only the upper corner of 
the 3$\sigma$ region allows $\geq 1$~$M_{\sun}$ neutron star mass. 
This suggests that their measurement is in general
too low to provide a realistic mass range for the neutron star.

On the basis of the current $K_{\rm comp}$ measurements,
$K_{\rm comp}\leq 410$ km~s$^{-1}$, since the K-correction factor is derived
assuming that $K_{\rm Bowen}$ indicated the orbital motion of the $L1$ point. 
Therefore from our fitting, $M_{\rm ns}\leq 1.3\ M_{\sun}$ at 1$\sigma$ level,
which confirms the point raised analytically by \citet{cor+09} 
and \citet{ele+09} that a light neutron star in \saxs\ is favored.

\subsection{Modulation in the Outburst}

While our short coverage of the outburst modulation prevents us from 
conclusive analysis, we have likely detected modulation changes during
the outburst. The second light curve appears to systematically deviate
from the simple sinusoidal-like shape, as seen in Figure~\ref{fig:lcr}.
This deviation could have been caused by disk flux variations during
the outburst, with a timescale of 20 min. On the other hand,
a modulation with a narrower flux peak can also explain this.
For example using our model, we qualitatively tested this and found 
that narrower light curves can indeed be produced with smaller irradiated 
areas, although much higher temperatures 
are required to account for similar maximum luminosity.
It is plausible that during the outburst, 
a hot spot could have formed on the surface of the companion due to strong 
irradiation by X-rays from the central pulsar.
\begin{center}
\includegraphics[scale=0.7]{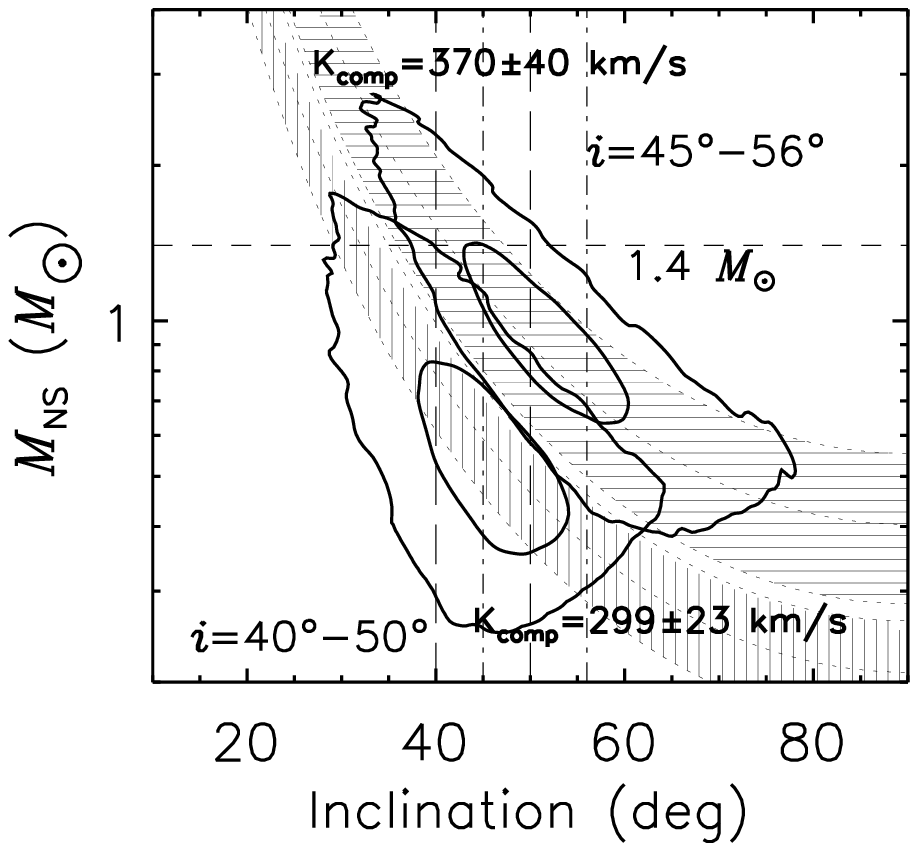}
\figcaption{$M_{\rm ns}$ vs. $i$ for $K_{\rm comp}=370\pm 40$ km s$^{-1}$
\citep{ele+09} and $K_{\rm comp}=299\pm 23$ km s$^{-1}$ \citep{cor+09}.
For the first and latter values, the 1- and 3-$\sigma$ constraints 
on $M_{\rm ns}$ and $i$, resulting from our fitting, are overplotted as
upper-right and lower-left contours, respectively. The derived 1-$\sigma$
$i$ value ranges are also indicated by vertical dash-dot and long dashed 
lines for the first and latter $K_{\rm comp}$ values, respectively.
\label{fig:mns}}
\end{center}

The third and fourth light curves,
3 days after the second one, changed to have a modulation with
0.1 mag amplitude. 
\citet{ele+09} have reported their detection of a slow-rise and a 
fast-decline modulation on 2008 Oct. 1, which would strongly
suggest the appearance of superhump modulation since or at least on 
Oct. 1 \citep{rlo97,wc10}. As observed in cataclysmic variables (CVs),
superhumps do often appear a few days after the start of an outburst 
(or superoutburst as in CVs). Before an accretion disk dumps nearly all 
its stored mass to the companion, the disk in the beginning of the outburst 
expands beyond the 3:1 resonance radius, which induces the tidal instability.
As a result, the disk develops into an eccentric form, producing superhump
modulation (e.g., see \citealt{osa96} for details). However in our data, 
not only are the light curves significantly different from that  
obtained on Oct. 1 by Elebert et al. (2009; for example,
its modulation amplitude was 0.2 mag), but also 
no clear phase shift of the modulation is seen. 
In the last two panels of Figure~\ref{fig:lcr}, we marked superior conjunction
of the companion (i.e., when the pulsar is at 270\arcdeg\ from the ascending
node) with two vertical dash-dotted lines, and as can be seen, the maximums
of the modulation were in phase with superior conjunction of the companion.
For superhump modulation in \saxs\ and given its mass ratio $q= 0.04/1.0$,  
the expected phase shift would be 0.1 per day, estimated from
the empirical relation between superhump and orbital periods and $q$ among 
known superhump binaries \citep{pat+05,wan+09}.

Our last two outburst light curves are similar to that obtained on
Oct. 7/8 by \citet{ele+09}. We note that there is marginal evidence 
that an additional component appears in the third light curve and that 
a weak trend exists in the residuals of the fourth light curve from 
our model fit.
Although the 0.1 mag amplitude modulation was in phase with that 
of the companion, we speculate that it
could consist of modulation components such as from the disk but
that modulation from the companion still dominated at the time. 
In order to understand this modulation,
better observational coverage of the source's outbursts and detailed 
analysis of each component's possible modulation should be needed. 
In any case, given that
the modulation showed significant variations over a few days, 
further observational studies of the source's modulation over
its future outbursts are warranted.

\subsection{Albedo of the Companion Star}

The day-side optical emission from the companion is due
to reprocessing of energy flux from the central pulsar. 
The required irradiation luminosities in quiescence in our model 
were in a range of (8--12)$\times 10^{33}$ erg~s$^{-1}$.
The known energy output from the pulsar in quiescence consists of its rotational
energy loss rate $L_{\rm sd}$ and X-ray emission, where the latter is 
two orders of
magnitude lower than the former. Therefore the energy output absorbed 
by the companion is approximately 
$(1-\eta_{\ast})f_b L_{\rm sd}\simeq 9\times 10^{33}(1-\eta_{\ast})f_b$ 
erg~s$^{-1}$, 
where $\eta_{\ast}$ is the albedo of the companion in quiescence and $f_b$ is
defined as the beaming factor for the pulsar's emission towards the companion.
Comparing this value to the irradiation luminosity range,
$\eta_{\ast}\simeq 0$ is required if $f_b\simeq 1$ 
(isotropic emission case). However from similar studies of
black widow pulsar binaries \citep{sta+01,rey+07}, $\eta_{\ast}$ was 
found to be around 0.6. If this value is taken for our case, $f_b\approx 2.5$
beaming for the pulsar's energy emission would be required.

Recently \citet{tct12} have proposed a model that irradiation 
could be due to $\gamma$-ray emission from pulsars for APMP
systems in quiescence. According to their calculations for \saxs, a high-mass 
($\sim$2 $M_{\sun}$) neutron star
is preferred considering that the irradiation luminosity 
is $L_{\gamma}\sim 10^{34}$ erg~s$^{-1}$.
Their model is therefore not fully supported by our derived limits on
the neutron star's mass but compatible within 3$\sigma$.

The required irradiation luminosity during the outburst on 2008 Sept. 29
was found to be 6.2$\times 10^{35}$ erg s$^{-1}$, while the data on
the other three nights are not considered because of the deviation of 
their modulations from a simple sinusoid-like shape. Nearly at the same time 
(MJD 54738.28), the 2.5--25 keV X-ray luminosity of the pulsar was 
3.2$\times 10^{36}$ erg s$^{-1}$ \citep{har+09}, where 4.1 kpc source distance 
(Table~\ref{tab:res}) is used.
Comparing the two values, the X-ray albedo of the companion 
$\eta_{\rm X}\simeq 0.8$ for the $f_b\simeq 1$ isotropic emission case. 
Such a large albedo value 
has been found in studies of accretion disks' X-ray reprocessing \citep{dva96}.
On the other hand, if $\eta_{\rm X}\simeq 0.6$ still holds, $f_b\simeq 1/2$
would be required.
The difference in the albedo or beaming factor during quiescence 
and the outburst probably provides additional evidence for having a 
different heating source in the quiescent state.

\section{Summary}
\label{sec:sum}

We have obtained nearly simultaneous $g'$ and $i'$ light curves of \saxs\  
in 2008 August and $r'$ light curves in 2008 September/October, respectively
before and after the reported 2008 September 22 X-ray outburst.
In the former datasets, we detected a clear
disk brightening as the optical precursor of the X-ray outburst.
In the latter datasets, the sinusoidal orbital flux modulation 
was observed to have an amplitude decrease
from 0.2 mag to 0.1 mag.

We employed an MCMC technique to fit all the quiescent data, which
include those previously published in \citet{del+08} and \citet{wan+09}, with
an irradiated companion model. We found a tight constraint on the inclination
angle of the binary system: $i=50^{+6}_{-5}$ deg. The resulting constraints
on the masses of the neutron star and companion were found to be 
0.75--1.28 $M_{\sun}$ and 0.03--0.06 $M_{\sun}$, 
respectively. These results rely on the $K_{\rm comp}$ measurement, which
currently is not certain, but (conservatively) considering
$K_{\rm comp}\leq 410$ km~s$^{-1}$, we found $M_{\rm ns}\leq 1.3\ M_{\sun}$
in \saxs.

From our fitting, the derived day-side temperature of the companion 
appeared to be constant over the period of $\sim$500 days 
during which
our quiescent data were obtained. The derived disk components can
be described by a simple disk temperature profile, while a change to a
steeper spectrum was possibly seen at the onset of the disk brightening. 

No modulation changes that might clearly reveal a different heating source 
during the outburst were found from our observations. 
Minor modulation variations, including a change of amplitude from 0.2 mag
to 0.1 mag, were detected in our outburst data.
We have possibly found
a narrower modulation peak than that seen in quiescence in one set of our data,
which could be
caused by the existence of a hot spot on the surface of the companion due
to intense irradiation by the X-ray pulsar. 
In order to further study these variations, good observational
coverage of the source during its outbursts is needed.


\acknowledgements

The Gemini queue mode observations were carried out under the program 
GS-2008B-Q-9. The Gemini Observatory is operated by the Association of 
Universities for Research in Astronomy, Inc., under a cooperative 
agreement with the NSF on behalf of the Gemini partnership: the National 
Science Foundation (United States), the Science and Technology 
Facilities Council (United Kingdom), the National Research Council 
(Canada), CONICYT (Chile), the Australian Research Council (Australia), 
CNPq (Brazil), and CONICET (Argentina). The CFHT observations were carried 
out under the TOO program 08BD97, obtained with MegaPrime/MegaCam, 
a joint project of CFHT and CEA/DAPNIA, at the Canada-France-Hawaii Telescope 
(CFHT) which is operated by the National Research Council (NRC) of Canada, 
the Institute National des Sciences de l'Univers of the Centre National 
de la Recherche Scientifique of France, and the University of Hawaii.

We thank Jacob M. Hartman for providing us the X-ray flux data during
the 2008 outburst of \saxs\ and anonymous referee for valuable
suggestions, and RPB thanks Rutger van Haasteren for discussions
about MCMC.
This research was supported by the starting funds of Shanghai
Astronomical Observatory, National Basic Research Program of China
(973 Project 2009CB824800), and 
National Natural Science Foundation of China (11073042).
ZW is a Research Fellow of the 
One-Hundred-Talents project of Chinese Academy of Sciences.
COH is supported by an Ingenuity New Faculty Award and an NSERC grant.

{\it Facility:} \facility{Gemini:South (GMOS), CFHT (MegaPrime/MegaCam)}

\bibliographystyle{apj}

\begin{deluxetable}{l c c c c c}
\tablecolumns{4}
\tablewidth{0pt}
\tablecaption{Fitted and Derived Model Parameters}
\tablehead{   
  \colhead{Parameter} &
  \colhead{Median} &
  \colhead{68.3\% interval} &
  \colhead{99.7\% interval} &
  \colhead{Reduced $\chi^2$\tablenotemark{a}} &
  \colhead{DoF}
}
\startdata
Inclination (degree) & 50 & 45--56 & 34--72 & \nodata & \nodata \\
$K_{\rm comp}$ (km\,s$^{-1}$) & 360 & 344--377 & 311--410  & \nodata & \nodata\\
Distance modulus & 13.08 & 12.97--13.19 & 12.74--13.41  & \nodata & \nodata\\
Absorption (J-band) & 0.26 & 0.20--0.32 & 0.09--0.42  & \nodata & \nodata\\
\tableline
\sidehead{2007 Mar 8 (MJD 54167.35; Gemini/GMOS-S)}
Day-side Temp. (K) & 8520 & 7870--9330 & 6960--11310 & \nodata & \nodata\\
Disk $g'$ ($\mu$Jy) & 7.4 & 5.0--10.8 & 0.9--27.1  & \nodata & \nodata\\
Noise & 2.3 & 1.8--2.9 & 1.2--5.0 & 1.0 & 40\\
Disk $i'$ ($\mu$Jy) & 15.1 & 11.1--20.1 & 2.3--38.9  & \nodata & \nodata\\
Noise\tablenotemark{b} & 19.0 & 9.6--46.8 & 1.7--441.3 & 0.74 & 4\\
\tableline
\sidehead{2007 Mar 10 (MJD 54169.35; Gemini/GMOS-S)}
Day-side Temp. (K) & 8110 & 7540--8800 & 6680--10540 & \nodata & \nodata\\
Disk $g'$ ($\mu$Jy) & 8.6 & 5.8--12.4 & 1.0--30.3  & \nodata & \nodata\\
Noise & 6.7 & 5.4--8.4 & 3.6--14.0 & 1.0 & 44\\
Disk $i'$ ($\mu$Jy) & 17.1 & 12.6--22.6 & 2.7--43.4  & \nodata & \nodata\\
Noise & 5.3 & 2.7--11.9 & 0.7--75.0 & 0.83 & 4\\
\tableline
\sidehead{2008 May 11 (MJD 54597.32; Gemini/GMOS-S)}
Day-side Temp. (K) & 8070 & 7440--8850 & 6530--10780  & \nodata & \nodata\\
Disk $r'$ ($\mu$Jy) & 11.2 & 8.0--15.4 & 1.7--32.5  & \nodata & \nodata\\
Noise & 32.5 & 25.6--41.5 & 16.6--74.1 & 0.97 & 36\\
\tableline
\sidehead{2008 May 12 (MJD 54598.34; Gemini/GMOS-S)}
Day-side Temp. (K) & 7990 & 7370--8740 & 6460--10540  & \nodata & \nodata\\
Disk $r'$ ($\mu$Jy) & 11.6 & 8.3--15.8 & 1.7--33.1  & \nodata & \nodata\\
Noise & 13.6 & 10.7--17.5 & 7.0--31.8 & 0.98 & 36\\
\tableline
\sidehead{2008 May 15 (MJD 54601.38; Gemini/GMOS-S)}
Day-side Temp. (K) & 8020 & 7400--8770 & 6510--10610  & \nodata & \nodata\\
Disk $r'$ ($\mu$Jy) & 12.6 & 9.0--17.2 & 1.8--36.2  & \nodata & \nodata\\
Noise & 12.7 & 10.0--16.3 & 6.4--29.9 & 1.0 & 36\\
\tableline
\sidehead{2008 Aug 4 (MJD 54682.08; Gemini/GMOS-S)}
Day-side Temp. (K) & 7720 & 7200--8310 & 6400--9650  & \nodata & \nodata\\
Disk $g'$ ($\mu$Jy) & 10.4 & 7.0--15.1 & 1.3--36.3  & \nodata & \nodata\\
Noise & 26.3 & 19.6--36.0 & 11.5--76.5 & 0.93 & 23\\
Disk $i'$ ($\mu$Jy) & 14.1 & 10.4--18.7 & 2.2--36.0  & \nodata & \nodata\\
Noise & 35.8 & 26.6--49.1 & 15.4--102.9 & 1.0 & 24\\
\tableline
\sidehead{2008 Aug 7 (MJD 54685.12; Gemini/GMOS-S)}
Day-side Temp. (K) & 7680 & 7150--8290 & 6350--9770  & \nodata & \nodata\\
Disk $g'$ ($\mu$Jy) & 37.2 & 25.1--53.6 & 4.8--127.3 & \nodata & \nodata\\
Noise & 21.4 & 17.4--26.6 & 12.0--43.1 & 0.96 & 47 \\
Disk $i'$ ($\mu$Jy) & 32.4 & 24.1--42.7 & 5.6--81.3  & \nodata & \nodata\\
Noise & 7.6 & 6.2--9.4 & 4.2--15.0 & 1.0 & 48\\
\cutinhead{Outburst data}
\sidehead{2008 Sept 29 (MJD 54738.21; CFHT/MegaCam)}
Day-side Temp. (K) & 22870 & 22200--23510 & 20740--24720 & \nodata & \nodata\\
Disk $r'$ ($\mu$Jy) & 292.3 & 291.6--293.0 & 290.2--295.3 & \nodata & \nodata\\
Noise & 1.6 & 1.3--1.9 & 0.9--2.9 & 0.94 & 60\\
\tableline
\sidehead{2008 Sept 30 (MJD 54739.21; CFHT/MegaCam)}
Day-side Temp. (K) & 24830 & 24560--24950 & 23630--25000 & \nodata & \nodata\\
Disk $r'$ ($\mu$Jy) & 292.6 & 292.4--293.0 & 292.3--294.8 & \nodata & \nodata\\
Noise & 3.2 & 2.6--3.9 & 1.9--5.9 & 1.0 & 57\\
\tableline
\sidehead{2008 Oct 3 (MJD 54742.21; CFHT/MegaCam)}
Day-side Temp. (K) & 12530 & 11960--13090 & 10820--14180 & \nodata & \nodata\\
Disk $r'$ ($\mu$Jy) & 240.4 & 239.6--241.2 & 237.0--243.9 & \nodata & \nodata\\
Noise & 1.7 & 1.5--2.0 & 1.1--2.8 & 1.0 & 84 \\
\sidehead{2008 Oct 4 (MJD 54743.21; CFHT/MegaCam)}
Disk $r'$ ($\mu$Jy) & 238.1 & 237.2--239.1 & 234.0--242.3 & \nodata & \nodata\\
Noise & 1.6 & 1.4--1.9 & 1.0--2.8 & 1.0 & 72\\
\cutinhead{Inferred Parameters}
Mass Comp. ($M_\odot$) & 0.04 & 0.03--0.06 & 0.02--0.11 & \nodata & \nodata \\
Mass NS ($M_\odot$) & 0.97 & 0.75--1.28 & 0.47--2.52  & \nodata & \nodata\\
\enddata
\label{tab:res}
\tablenotetext{a}{The quoted reduced-$\chi^2$ were calculated at the median
parameter values, and are provided for indicative purposes only since they 
do not take into account the priors nor full Bayesian posterior distributions.}
\tablenotetext{b}{Noise parameter $b$ (see \ref{sec:noise})}
\end{deluxetable}

\end{document}